\documentclass[aps,prl,reprint,superscriptaddress,showkeys,amsmath,amssymb,floatfix]{revtex4-2}

\usepackage{amsfonts}
\usepackage{siunitx}
\usepackage{setspace}
\usepackage{hyperref}
\usepackage{orcidlink}

\renewcommand{\Im}{\textup{Im}}

\begin{document}

\bibliographystyle{apsrev4-2}

\title{Reply to ``Comment on `Gravitational Pair Production and Black Hole Evaporation'{}''}

\author{Michael F. Wondrak\,\orcidlink{0000-0002-6894-1072}}
\email[]{m.wondrak@astro.ru.nl}
\affiliation{Department of Astrophysics/IMAPP, Radboud University, P.O.\ Box 9010,
6500 GL Nijmegen, The Netherlands}
\affiliation{Department of Mathematics/IMAPP, Radboud University, P.O.\ Box 9010,
6500 GL Nijmegen, The Netherlands}

\author{Walter D. van Suijlekom\,\orcidlink{0000-0003-4507-5041}}
\email[]{waltervs@math.ru.nl}
\affiliation{Department of Mathematics/IMAPP, Radboud University, P.O.\ Box 9010,
6500 GL Nijmegen, The Netherlands}

\author{Heino Falcke\,\orcidlink{0000-0002-2526-6724}}
\email[]{h.falcke@astro.ru.nl}
\affiliation{Department of Astrophysics/IMAPP, Radboud University, P.O.\ Box 9010,
6500 GL Nijmegen, The Netherlands}

\date{18 April 2024}

\begin{abstract}%
In a recent letter, the authors presented a unified derivation of the electric Schwinger effect and a generalized Hawking effect with an additional radiation component. The approach discloses a radial profile of black hole pair production and traces the emission back to local tidal forces which are independent of the black hole event horizon. It uses an effective action valid to second order in curvature and arbitrary order in proper time.
A comment on the letter supposed two inconsistencies when applying the central formula to the Schwinger effect in the presence of magnetic fields. 
The present letter points out that the partially flawed argumentation does not cast doubt on the results. 
\end{abstract}


\maketitle

A recent letter~\cite{Wondrak:2023zdi} by the authors showed a new avenue to particle production in the spacetime of a Schwarzschild black hole. 
This approach based on covariant perturbation theory~\cite{Barvinsky:1990up,Codello:2012kq} predicts the creation of scalar particles from an imaginary part of the effective action.  
Comment~\cite{Ferreiro:2023jfsWOarXiv} raises four questions regarding the central formula.
This letter points out that the comments are based on an incorrect interpretation, go beyond the range of applicability of the formula, or have already been discussed.

The first comment erroneously assumes that the expression for the effective action is a lowest-order approximation to a series in proper time $s$. However, the principle of covariant perturbation theory is an expansion in powers of curvature. As evident from Appendix S.1 of the original letter~\cite{Wondrak:2023zdi}, a truncation to any finite order in $s$ will lead to the same expression.

The second comment refers to the Schwinger effect in the presence of magnetic fields. In contrast to the gradients in the electric or in the Schwarzschild case, magnetic fields introduce more complex paths of the virtual field excitations responsible for particle production. 
The central formula of the original letter~\cite{Wondrak:2023zdi} considers terms up to second order in curvature, including the term 
$\Omega_{\mu\nu} \Omega^{\mu\nu}$ corresponding to one of the electromagnetic invariants, 
$\mathcal{F} = F_{\mu\nu} F^{\mu\nu}/4 = (\vec{B}^2 -\vec{E}^2)/2$ 
\footnote{%
The curvature of the electromagnetic connection, $\Omega_{\mu\nu}$, is proportional to the Faraday field-strength tensor $F_{\mu\nu}$ as $\Omega_{\mu\nu} = \text{i} q F_{\mu\nu}/\hbar$.
Note that we follow the $(-{}+{}+{}+)$ signature convention for the metric, set $c \equiv 1$, and define the totally antisymmetric symbol by $\epsilon_{0123} = 1$.%
}. 
This formula is capable to correctly reproduce the electric Schwinger effect.

Only for non-zero magnetic fields, the other electromagnetic invariant,
$\mathcal{G} = F_{\mu\nu} \tilde{F}^{\mu\nu}/4 = \vec{E} \cdot \vec{B}$, needs to be considered. 
Being an axial quantity, however, one could expect $\mathcal{G}$ to enter the effective action starting at fourth order in curvature~\cite[p.~205]{ParkerToms2009} 
\footnote{%
Note that $\mathcal{F}$ and $\mathcal{G}$ are quadratic in the field strengths so that $\mathcal{G}$ is a second-order term and $\mathcal{G}^2$ is a fourth-order term in the curvature expansion.%
}
and therefore outside the range of applicability of the central formula in~\cite{Wondrak:2023zdi}.

The comment~\cite{Ferreiro:2023jfsWOarXiv}, however, erroneously states that the term
$\vec{E} \cdot \vec{B}$, i.e.\ the invariant $\mathcal{G}$, would be the leading-order contribution in the curvature expansion.
The argumentation is based on the closed-form expression for the imaginary part of the effective Lagrangian in the case of parallel electric and magnetic fields, given by~\cite{Popov:1971iga,Dunne:2004nc}.

Expressing this imaginary part in terms of $\mathcal{F}$ and $\mathcal{G}$ (thereby also applicable to general non-orthogonal field orientations~\cite{Popov:1971iga}), 
\begin{align}
\label{eq:ImLeffEBITOInvariants}
\Im\! \left(\mathcal{L}_\text{eff}\right)
\sim \frac{q^2\, \mathcal{G}}{16 \pi^2\, \hbar}\, 
  \sum_{n=1}^\infty 
    \frac{{\left(-1\right)}^{n-1}}{n\, \sinh\!\left(\frac{\mathcal{F} +\sqrt{\mathcal{F}^2 +\mathcal{G}^2}}{\mathcal{G}}\, n\pi\right)},
\end{align}
one can show the contrary in the two relevant regimes:\\
A) For a weak magnetic field, the invariant enters in even powers, i.e.\ 
$\mathcal{G}^2 = (\vec{E} \cdot \vec{B})^2$ and higher,
\begin{align}
\Im\! \left(\mathcal{L}_\text{eff}\right)
\sim -\frac{q^2\, \mathcal{F}}{16 \pi^2\, \hbar} \left(
  c_0 +c_2\, \frac{{\mathcal{G}}^2}{{\mathcal{F}}^2} 
  +\ldots
\right),
\end{align}
with $c_0 \approx \num{0.524}$ and $c_2 \approx \num{0.131}$.\\
B) For equally strong fields, 
$|\vec{E}| = |\vec{B}|$, 
i.e.\ $\mathcal{F}=0$, 
the dependency on the electromagnetic fields reduces to a factor 
$\left|\mathcal{G}\right|$, 
\begin{align}
\Im\! \left(\mathcal{L}_\text{eff}\right)
&\sim \frac{q^2\, \left|\mathcal{G}\right|}{16 \pi^2\, \hbar}\, 
  \sum_{n=1}^\infty 
    \frac{{\left(-1\right)}^{n-1}}{n\, \sinh\!\left(n\pi\right)}. 
\end{align}
Also the correction terms around $\mathcal{F}=0$ depend on $\mathcal{G}$ only via the absolute value.
Since the absolute value is a non-analytic function, the term $\left|\mathcal{G}\right|$  could arise from higher-order terms in the context of resummation, but it cannot occur as the leading term in a curvature expansion.

The third comment concerns a potential unitarity violation.
The specific difference between a purely electric and magnetic field has already been discussed in~\cite{Chernodub:2023pwf} where unitarity is ensured by granting non-negative production probabilities. 
In the fermionic example, going beyond the scalar case in the letter, the only non-zero electromagnetic invariant is $\mathcal{G}$. However, considering the spinor case~\cite{Dunne:2004nc} in analogy to the discussion in the preceding paragraphs, 
$\mathcal{G}$ would only give positive contributions to the imaginary part of the effective action. Therefore, no unitarity violation is expected.

The fourth comment addresses a general topic already discussed in the original letter~\cite{Wondrak:2023zdi}. The relation between the presence of particle production and the choice of the quantum vacuum state is also an ongoing debate for Hawking radiation, e.g.\ for Boulware-like states~\cite{Mathur:2023uoe,Mathur:2024mvo}. 
While there might be state-dependent contributions beyond the current order in the curvature expansion, we highlight the consistency of the apparent state-independency with the conformal anomaly which is also state-independent~\cite{Chernodub:2023pwf}.
Furthermore, particle-production formulae similar to the expression for the imaginary part of the effective action have been found e.g.\ in the case of time-dependent anisotropies~\cite{Zeldovich:1977vgo}.

The main findings in the original letter line up with a tradition of ideas in literature. For instance, it has been demonstrated that event horizons are neither necessary for particle production in general~\cite{Hossenfelder:2002ag} nor for the Hawking effect for time-dependent~\cite{Visser:2001kq} and static~\cite{Mathur:2023uoe,Mathur:2024mvo} backgrounds. 
Additionally, the conformal anomaly is known as an approach to describing particle production by black holes~\cite{Christensen:1977jc}.
Moreover, an extended particle emission region around a black hole has also been derived in~\cite{Giddings:2015uzr,Dey:2017yez}. The magnitude of the results in the original letter fits well to these publications.

In summary, we do not share the concern of comment~\cite{Ferreiro:2023jfsWOarXiv} about the main findings of the original letter~\cite{Wondrak:2023zdi}. 
The argument based on a concrete example is incorrect. 
The other considerations do not cast doubt on the results presented in letter~\cite{Wondrak:2023zdi} because they have already been addressed or lie outside the realm of applicability. In contrast, they strengthen the approach as one can demonstrate where and why they fall short.

\begin{acknowledgments}
The authors thank M.\ Chernodub for fruitful discussions.
This work was supported by an Excellence Fellowship from Radboud University and a grant from NWO NWA 6201348. 
\end{acknowledgments}

\end{document}